\documentclass[reprint,amsmath,amssymb,aps,showpacs,superscriptaddress,twocolumn,prl]{revtex4-1}
\usepackage{graphicx}
\usepackage{bm}
\usepackage{epsfig}
\usepackage{amssymb}
\usepackage{amsfonts}
\usepackage{braket}
\usepackage{color}
\usepackage{epstopdf}
\usepackage{hyperref}
\usepackage{float}
\restylefloat{table}
\usepackage{bibentry}
\usepackage{multirow}
\usepackage[caption=false]{subfig}

\graphicspath{{figures/}}% Put all figures in this directory.

\newcommand{\ba}{\begin{eqnarray}}
\newcommand{\ea}{\end{eqnarray}}
\newcommand{\bd}{\begin{displaymath}}

\begin{document}
\title{ Spin-1 bilinear-biquadratic model on star lattice }
\author{Hyun-Yong Lee}
\email{hyunyong.rhee@gmail.com}
\affiliation{Institute for Solid State Physics, University of Tokyo, Kashiwa, Chiba 277-8581, Japan}
\date{\today}

\author{Naoki Kawashima}
\email{kawashima@issp.u-tokyo.ac.jp}
\affiliation{Institute for Solid State Physics, University of Tokyo, Kashiwa, Chiba 277-8581, Japan}
\date{\today}

\begin{abstract}
 We study the ground-state phase diagram of the $S=1$ bilinear-biquadratic model\,(BLBQ) on the star lattice with the state-of-art tensor network algorithms. The system has five phases: the ferromagnetic, anti-ferromagnetic, ferroquadrupolar, and spin-liquid phases. The phases and their phase boundaries are determined by examining various local observables, correlation functions and transfer matrices exhaustively. The spin liquid phase, which is the first quantum disordered phase found in two-dimensional BLBQ model, is gapped and devoid of any conventional long-range order.  It is also characterized by fixed-parity virtual bonds in the tensor network formalism, analogous to the Haldane phase, while the parity varies depending on the location of the bond.
 \end{abstract}
\maketitle

{\it Introduction-} After the discovery of the high-temperature superconductors\,\cite{bednorz86}, their parent compounds are conjectured to be in a spin liquid\,(SL) phase which becomes superconducting when charge carriers are doped\,\cite{anderson87}. 
%SL refers to a ground state\,(GS) where strong quantum fluctuations frustrate a conventional magnetic ordering and let the spins stay in a disordered state without breaking any symmetry down to absolute zero temperature\,\cite{balents10}. 
Such SLs are expected to possess a kind of quantum order\,\cite{wen02, kitaev06, han12}, e.g. $Z_2$ topological order in $Z_2$-SLs\,\cite{wen02, kitaev06}, or support fractionalized edge excitations protected by some symmetries\,\cite{wen14}, e.g. Haldane phase\,\cite{haldane83, aklt88, gu09, pollmann12}. Quantum effects or fluctuations are believed to become stronger as the spin and spatial dimension decrease. The geometric frustration also plays an important role\,\cite{balents10}. Consequently, with the successful realizations of the kagome lattice in volborthite\,\cite{zenji01}, herbertsmithite\,\cite{shores05} and kapellasite\,\cite{hiroyuki17}, and the triangular lattice in $\kappa$-BEDT(CN)$_3$\,\cite{shimizu03}, the frustrated $S=\frac{1}{2}$ systems have been extensively studied to find stable SL states. However, the recent discoveries of SLs in the pnictide family of superconductors\,\cite{qazilbash09, yin11}, and an unconventional quantum disordered state in Ba$_3$NiSb$_2$O$_9$\,\cite{cheng11} triggered a burst of investigations on $S=1$ quantum magnets on square\,\cite{harada02, harada07, toth12, wang15, cheng16, hylee16, hylee17, gong17, ido17a, ido17b}	, honeycomb\,\cite{lee12, tao12}, kagome\,\cite{liu15} and triangular\,\cite{lauchli06a, serbyn11} lattices, respectively. Theoretically, such lattices can be decorated to be a so-called ``star lattice" of which the geometry is distinct from the ones of lattices listed above. Generally, such decoration may cause non-trivial results on the state with strong fluctuations. One may, therefore, seek novel spin liquid states in such lattices\,\cite{barabanov94, canals02, corboz10}. In fact, previous studies on the star lattice spin-$1/2$ models found an exact chiral SL with non-Abelian anyonic excitations\,\cite{yao07, kells10}, various valence-bond-solid\,(VBS) ground states\,(GS)\,\cite{ybj10} and topological order in several SL phases\,\cite{huang13}. 
%However, the $S=1$ spin models on the star lattice are much more challenging than the models on the commonly studied lattices due to the large unit-cell, and we are not aware of any publication so far.

\begin{figure}[!b]
  \includegraphics[width=0.5\textwidth]{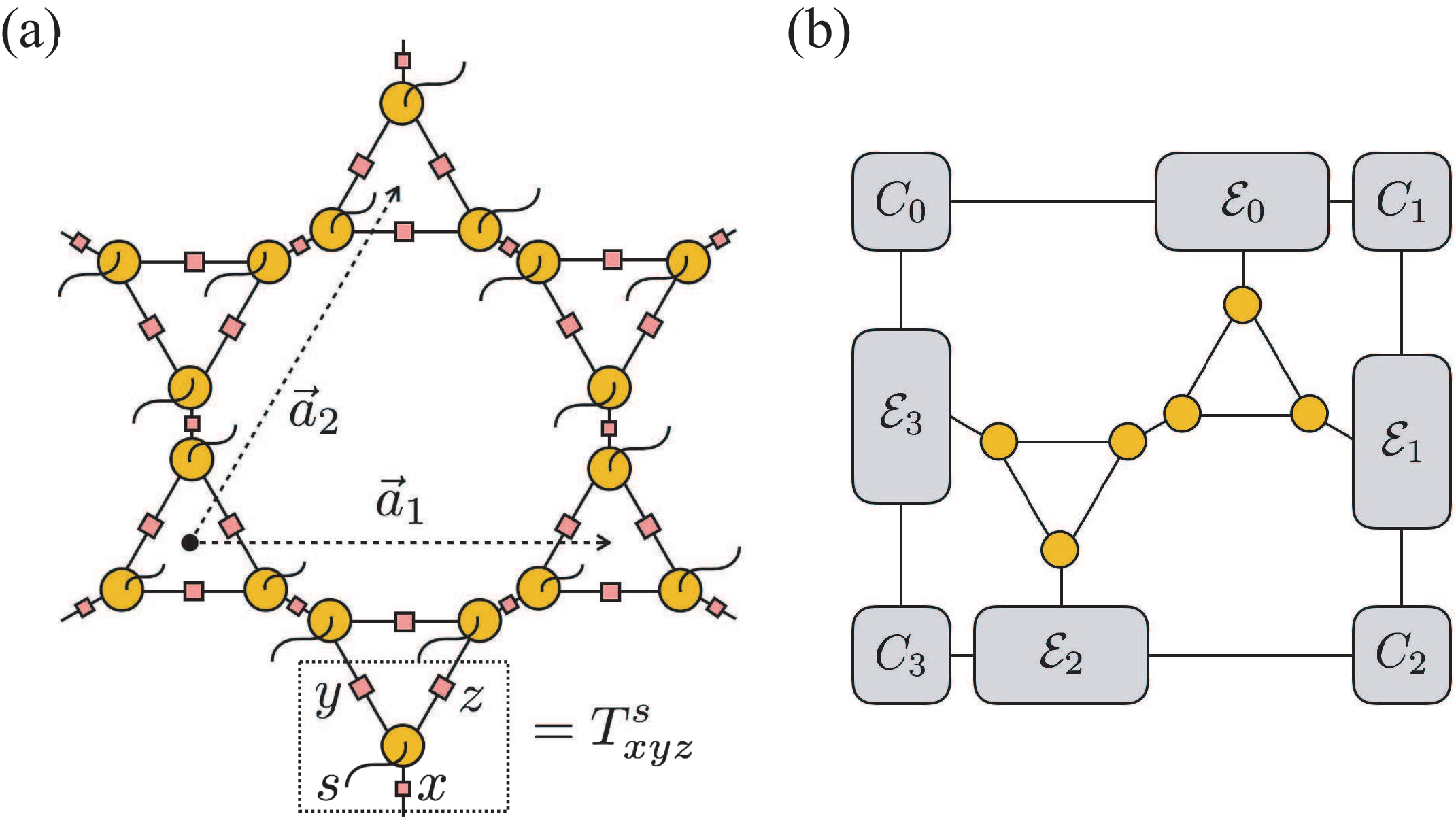}
  \caption{  Schematic figures of (a) iPEPS\,(yellow circle: site tensor, red square: singular value matrix) on the star lattice\,($\vec{a}_{1,2}$: lattice vector) and (b) environment tensors\,$\{ C_i, \mathcal{E}_i \}$ and double layered tensors in the unit-cell. Here, $T_{xyz}^s$ in the dashed box denotes the site tensor\,(see the main text for details).}
  \label{fig:schematic}
\end{figure}

Theoretical and computational studies on the strongly correlated systems are entering a new phase under the remarkable development in the tensor network\,(TN) algorithms. We refer the readers to Ref.\,\onlinecite{orus14} for an exhaustive list of relevant literatures. The TN method does not suffer from the sign problem for the frustrated models and also allows us to reach the thermodynamic limit efficiently by employing the framework of renormalization group\,\cite{levin07,gu08}. Advantage of the tensor network representation is not only technical but also conceptual; information on the GS entanglement is directly accessible by looking at the geometry of TN and gauge symmetry of local tensors\,\cite{orus14,shenghan15}. A well-known example is the Haldane phase (and its generalization to higher dimensions) that can be characterized very clearly by the fixed parity of the virtual bonds in their tensor network representations\,\cite{pollmann12}. In this sense, the TN method is ideal to investigate SL, and a lot of approaches have been already proposed in recent years\,[see Ref.\,\onlinecite{mei17} and references therein]. In the present Letter, we employ TN algorithms to explore the $S=1$ BLBQ model on the star lattice\,[Fig.\,\ref{fig:schematic}\,(a)].

{\it Model-} Let us begin with defining the BLBQ model:
\begin{align}
	H 
	%&= \sum_{\langle i,j\rangle}\left[ \cos\phi ({\bm S}_i\cdot {\bm S}_j) + \sin\phi ( {\bm S}_i\cdot {\bm S}_j )^2 \right]\nn
	& = \sum_{\langle i,j\rangle} \left[ \left( \cos\phi - \frac{\sin\phi}{2} \right){\bm S}_i\cdot {\bm S}_j + \frac{\sin\phi}{2} {\bm Q}_i \cdot {\bm Q}_j \right]
	%& = \sum_{\langle i,j\rangle} (\cos\phi-\sin\phi) ({\bm S}_i\cdot {\bm S}_j) + \sin\phi(1+P_{ij})
	\label{eq:ham}
\end{align}
where $\langle i,j\rangle$ denotes the nearest neighbor sites, ${\bm S}_i$ the spin-1 operator and ${\bm Q}_i$ the quadrupolar\,(QD) operator with 5 components: $(S_i^x)^2-(S_i^y)^2,\,\sqrt{3}(S_i^z)^2-2/\sqrt{3},\, S_i^x S_i^y + S_i^y S_i^x,\, S_i^y S_i^z + S_i^z S_i^y$ and $S_i^z S_i^x + S_i^x S_i^z$. The quadrupole moment $Q \equiv N_s^{-1} \sum_i^{N_s} \sqrt{ \langle {\bm Q}_i \rangle^2}$, where $N_s$ is the total number of lattice sites, is a fundamental order parameter identifying the phases in the BLBQ model on various lattices\,\cite{schmitt98, harada02, lauchli06a, lauchli06b, harada07, tao12, wang16}. Note that at some special values of $\phi$ the system possesses the symmetry higher than the obvious $SU(2)$ symmetry. At $\phi=-3\pi/4,\pi/4$, the Hamiltonian is invariant under simultaneous $SU(3)$ rotations at all lattice points. 

%In addition, on bipartite lattices, the system has the $SU(3)$ symmetry also at $\phi=\pm\pi/2$. Namely, if we define a local operator $W$ that swaps $S_z=\pm 1$ states and changes signs of those components, the Hamiltonian is invariant under $SU(3)$ rotation $U$ applied to all lattice points on one sublattice, and $WUW$ to the other sublattice. In the present case, however, due to the geometrical frustration, the system does not enjoy this full $SU(3)$ symmetry. 
 
{\it Method-} To carve out the GS phase diagram of the model in Eq.\,\eqref{eq:ham}, we optimize the inifinite projected entangled pair states\,(iPEPS) with a rank-4 site tensor $T_{x_i y_i z_i}^{s_i}$ and singular value matrices $\lambda_{\alpha_i \alpha_i'}$ \,\cite{verstraete08,tao08}:
%
%\begin{align}
$	|\psi\rangle = \sum_{\{\alpha_i\}}{\rm Tr} \prod_i T_{x_i y_i z_i}^{s_i} \lambda_{x_i x_i'}^{1/2}\lambda_{y_i y_i'}^{1/2}\lambda_{z_i z_i'}^{1/2}|s_i\rangle,
$
%\end{align}
%
where ${\rm Tr}$ represents the trace over the virtual indices\,($x_i,y_i,z_i$), and $s_i$ is the local quantum number. Its graphical representation is shown in Fig.\,\ref{fig:schematic}\,(a). For convenience later, we define the bond connecting two triangle plaquettes as $x$-bond and two bonds forming a triangle plaquette as $y$- and $z$-bond in the clockwise direction as depicted in Fig.\,\ref{fig:schematic}\,(a). By applying iteratively the imaginary-time evolution operator\,[$\exp(-\tau H_{ij})$] on every bond, one can optimize $T_{x_i y_i z_i}^{s_i}$ with respect to the energy density. The simple update\,(SU) is a popular method to renew the tensors at every imaginary-time step\,\cite{tao08}. Recently, the importance of preserving symmetries in optimization has been noticed with a development of so-called ``symmetric" simple update\,(SSU), which allows us to keep symmetries throughout the imaginary-time evolution\,\cite{mei17}. In this paper, either SU or SSU is adopted depending on the initial and target states. To be more precise, we examine three kinds of ansatz: $SU(2)$ symmetric, time-reversal\,(TR) symmetric and non-constraint ansatz. We try several initial conditions for each type of ansataz, e.g., the ferromagnetic\,(FM), $120^{\circ}$ coplanar antiferromagnetic\,(AFM) product states and random states for the non-constraint ansatz.
%Interestingly, we found that SSU improves the optimization and thus leads to a (energetically) better ansatz than the ones obtained by SU starting from random initial states, which is not guaranteed in the general case, over a finite-range of parameter $\phi$. 
In order to contract iPEPS without symmetry breaking, we apply the basic idea of SSU to the corner transfer matrix renormalization group\,(CTMRG) method\,\cite{baxter68, nishino96, nishino98, orus09, corboz10, orus12}. Then, we measure the physical quantities, such as the local order parameters and correlation functions, using the environment tensors obtained by CTMRG. The parallel C++ library {\it mptensor}\,\cite{mptensor} is utilized to perform all TN algorithms in the present work.

\begin{figure}[!t]
  \includegraphics[width=0.3\textwidth]{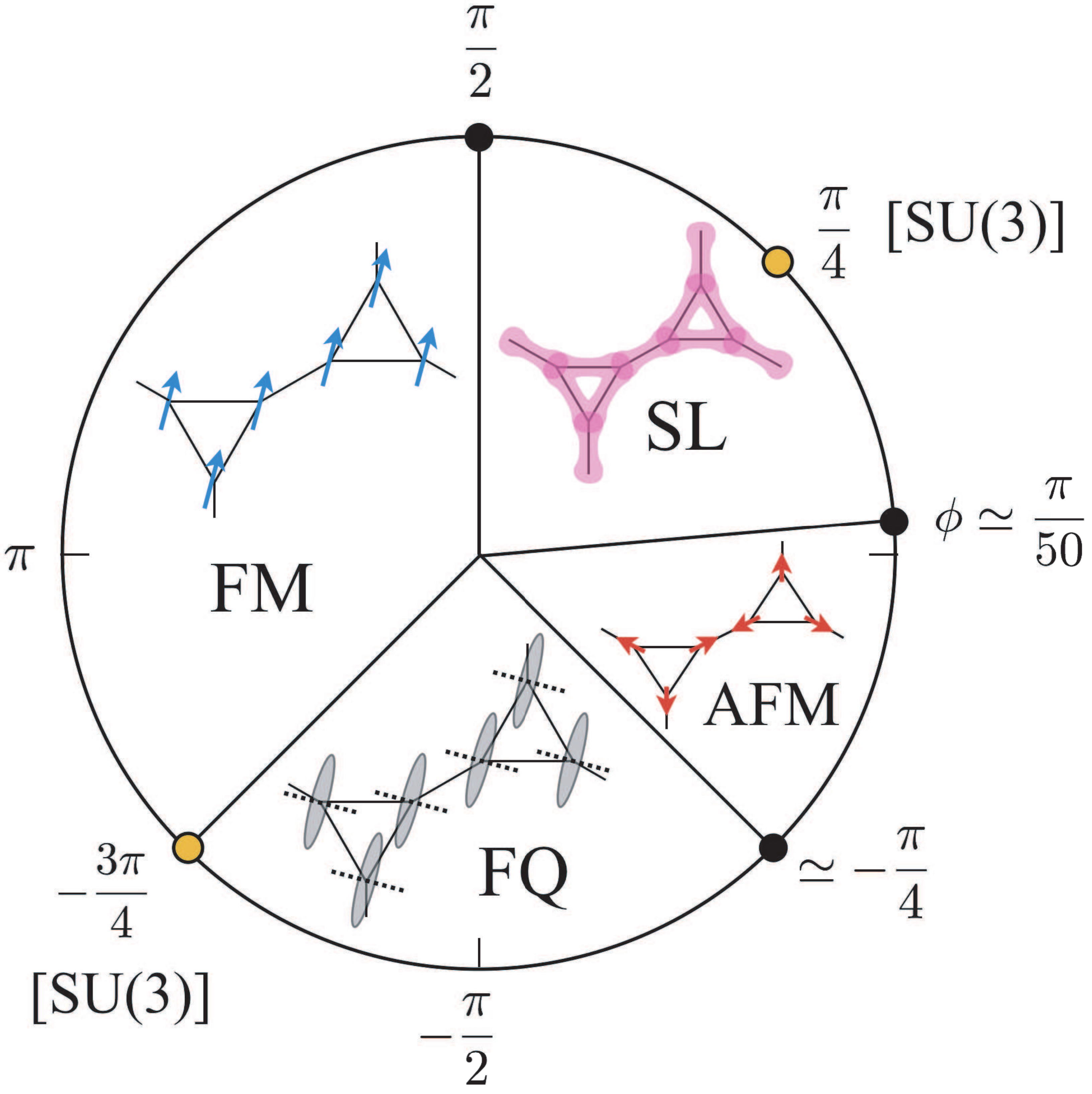}
  \caption{Phase diagram of the BLBQ model on the star lattice\,[Eq.\,\eqref{eq:ham}] as a function of the mixing angle $\phi$. Here, FM, FQ, AFM and SL represent ferromagnetic, ferroqudrupolar, $120^{\circ}$ coplanar antiferromagnetic and spin liquid phases, respectively. The model has the $SU(3)$ symmetry at $\phi = -0.75\pi, 0.25\pi$ which are denoted by yellow circles.}
  \label{fig:phase_diagram}
\end{figure}

\begin{figure}[!b]
  \includegraphics[width=0.5\textwidth]{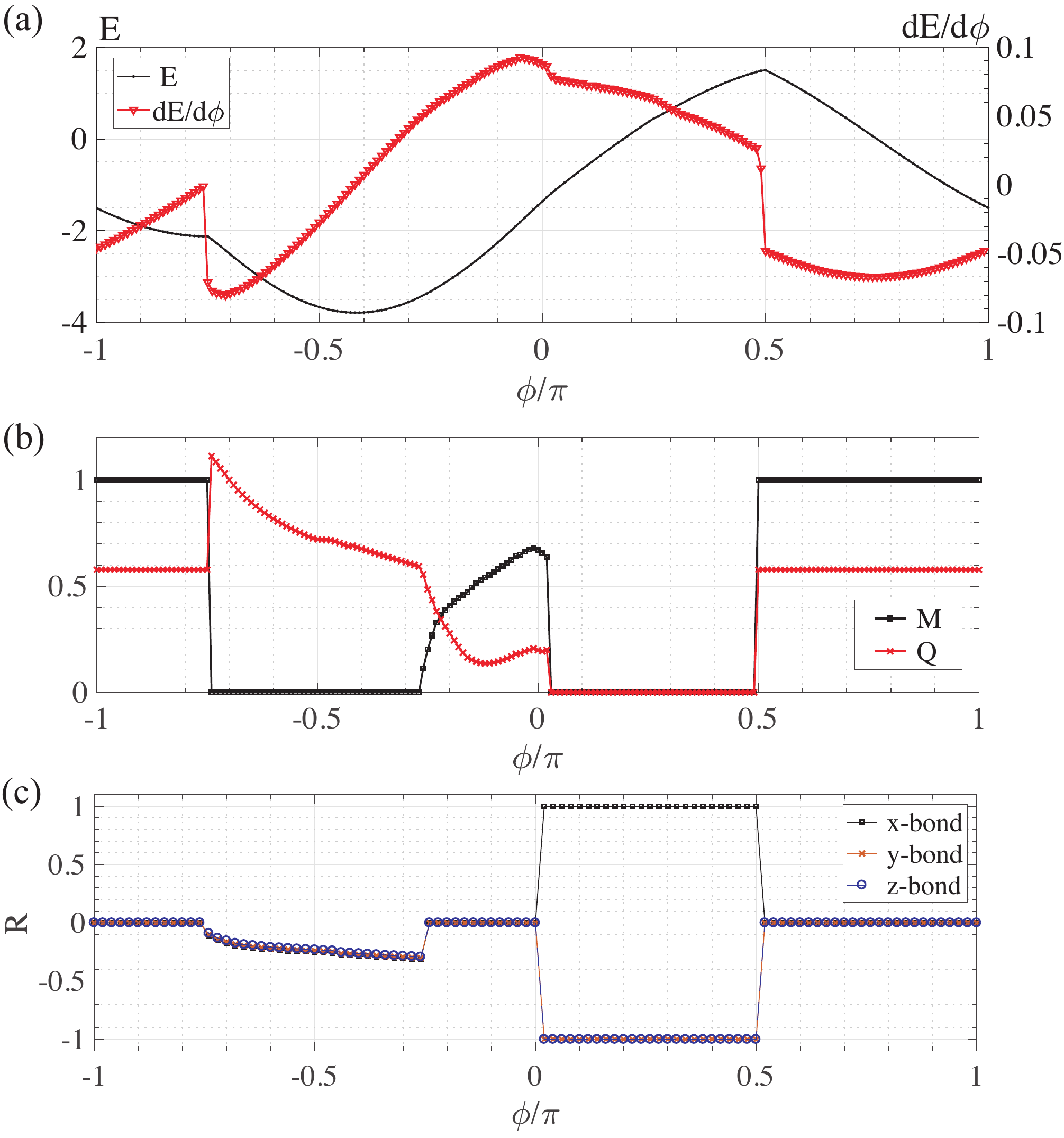}
  \caption{ Plots of the (a) energy density, (b) magnetization and quadrupole moments of GS wave function and (c) the quantity $R$ defined in Eq.\,\eqref{eq:R} as a function of $\phi$, respectively. }
  \label{fig:EMQ}
\end{figure}

{\it Identification of each phase-} GS phase diagram is presented in Fig.\,\ref{fig:phase_diagram}, in which five phases are identified: FM, ferroquadrupolar\,(FQ), AFM and a SL phase. We have determined those phases by analyzing the energy density, local order parameters and the connected correlation functions for the optimized ansatz on a variety of unit-cell structures\,\cite{corboz11} with trial initial states. The bond dimension $D$ is varied from 1 to 12, and the GSs are adopted at each $\phi$ by the lowest energy density shown in Fig.\,\ref{fig:EMQ}\, (a). Here, we identify and discuss the properties of each phase, and then the nature of phase boundaries will be discussed afterwards.

%

%{\it FM phase-} 
Regardless of the spatial dimension or the lattice geometry, the BLBQ model exhibits FM phase in $0.5\pi<\phi<1.25\pi$\,\cite{lauchli06b, pjkim16, lauchli06a, tao08, liu15} as we also found. In this phase, the imaginary time evolution leads the tensors to a trivial tensor with $D=1$, i.e. a product state. Throughout this phase, the magnetization $M \equiv N_s^{-1} \sum_i^{N_s} \sqrt{ \langle {\bm S}_i \rangle^2}$ is always 1, and $Q=1/\sqrt{3}$ indicating fully aligned spins. 

%{\it FQ phase-} 
The FQ phase occurs right next to FM phase at $\phi=-0.75\pi$ and disappears at $\phi\simeq-0.25\pi$ where the GS enters into AFM phase. Since BQ interaction with negative sign favors parallel alignment of the quadrupole moments, the FQ state becomes stable immediately after BQ wins BL exchange\,($\phi>-0.75\pi$). The FQ order parameter gradually decreases from the largest value $Q(\phi=-0.75\pi)=2/\sqrt{3}$ as $\phi$ approaches to $-0.25\pi$, while the magnetization is always zero up to the machine precision. The TR-symmetric initial state with SSU flows into the lowest energy state resulting in $M=0$ all through this phase. We find that $\lambda_{\alpha_i \alpha_i'}$ are rotationally symmetric\,(i.e., $\lambda_{x_i x_i'}=\lambda_{y_i y_i'}=\lambda_{z_i z_i'}$) and carry non-degenerate and doubly degenerate values. It denotes that the site tensor accommodates the Kramers singlets and doublets on the virtual legs to form a TR-symmetric tensor.

%{\it AFM phase-} 
As $\phi$ passes through $-0.25\pi$, the magnetization gradually increases from zero\,[Fig.\,\ref{fig:EMQ}\,(b)], and spins form the $120^{\circ}$ coplanar configuration. The FQ order parameter remains finite due to the finite magnetization. The magnetization reaches the maximum at $\phi=0$ where BQ exchange is turned off, and this is similar to the triangular and honeycomb models\,\cite{lauchli06a, tao12}. AFM phase seems to extend to $\phi\simeq 0.02 \pi$. However, it is not exactly determined as the iPEPS optimization does not converge well and thus shows some fluctuations in the energy density and order parameters over $0<\phi\lesssim 0.02\pi$. Nevertheless AFM state gives still the lowest energy than others.

\begin{figure}[!t]
  \includegraphics[width=0.5\textwidth]{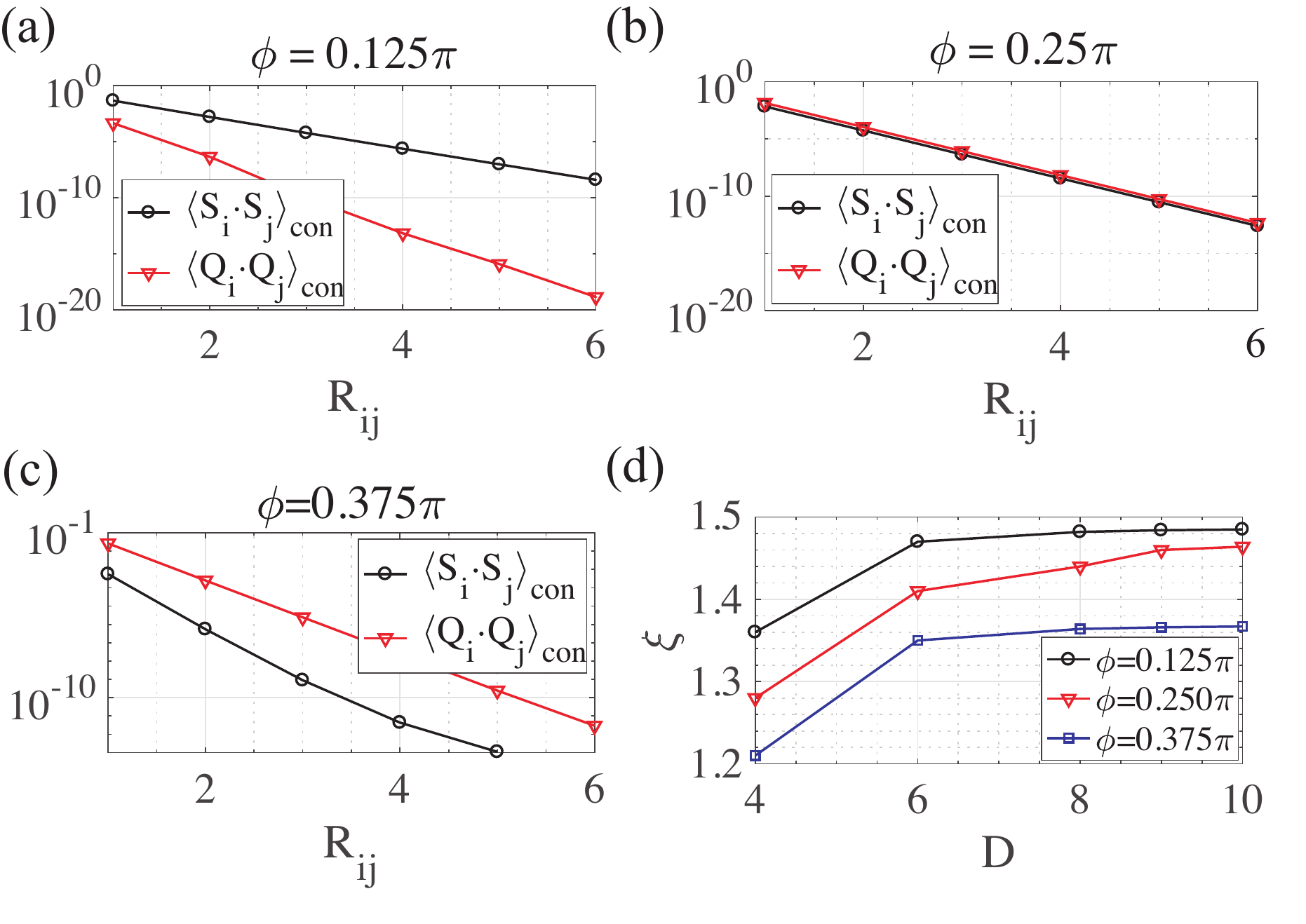}
  \caption{Spin-spin and quadrupole-quadrupole correlations at (a) $\phi=0.125$ and (b) $\phi=0.375\pi$, respectively. Here, $R_{ij}$ is the distance between $i$- and $j$-th sites in units of $|\vec{a}_1|$ defined in Fig.\,\ref{fig:schematic}.}
  \label{fig:correlations}
\end{figure}
%

%{\it SL phases-}
%Two distinct SL phases appear in $0.02\pi\lesssim \phi < 0.5\pi$, and those are separated at $\phi=0.5\pi$ where the symmetry is enhanced to $SU(3)$. In fact, the emergence of SL phases is quite surprising in that only symmetry broken phases are found between AFM and FM phases such as plaquette-VBS on the honeycomb\,\cite{tao12} and anti-ferroquadrupolar on the triangular lattice\,\cite{lauchli06a}. In this phase, the $SU(2)$-symmetric ansatz with SSU on $(1\times1)$ unit-cell provides the GS wavefunction. 

In $0.02\pi\lesssim \phi < 0.5\pi$, the $SU(2)$-symmetric SSU on $(1\times1)$ unit-cell provides the best ansatz, and therefore the GS is SL throughout this region. 
%To our best knowledge, this is the first quantum disordered phase reported in two-dimensional BLBQ model\,\cite{lauchli06a,tao12,liu15}. 
By virtue of SSU, we find that only the integer spins are accommodated on the $x$-bond while only the half-integer spins on the $y$- and $z$- bonds. In order to show this interesting feature, we define a quantity
\begin{align}
	R \equiv \frac{\sum_i (d_i-1)(-1)^{d_i-1} \lambda_i}{\sum_i \lambda_i},
	\label{eq:R}
\end{align}
where $d_i$ is the degeneracy of the $i$-th singular value $\lambda_i$. The $R$ at each bond is presented in Fig.\,\ref{fig:EMQ}\,(c). Due to lack of any symmetry, the $\lambda_i$ does not degenerate, and therefore $R$ is zero throughout the FM and AFM phases. It is finite and changes continuously in the FQ phase because of the double degeneracy of some of the singular values guaranteed by TR-symmetry. In the SL phase, the $R$ becomes integer either $+1$ or $-1$ depending on the bond, which shows that each bond accommodates either only even- or odd-parity multiplets. This is analogous to the Haldane phase where the virtual bonds carry only the odd-parity multiplets\,\cite{pollmann12}.

The positive BQ exchange favors a perpendicular orientation of neighboring quadrupole moments. It induces the anti-ferroquadrupolar phases on the triangular\,\cite{lauchli06a} and kagome\,\cite{liu15} lattices, on which spins on a triangle plaquette are shared by neighboring plaquettes. Instead, on the star lattice,  spins are not shared but entangled with others on neighboring plaquttes. As a result, it may give rise to configurational fluctuations of quadrupole moments and subsequently the recovery of symmetry. 

In order to explore the physical property of the SL phase, we evaluate the connected spin $C_{ij}^S\equiv\langle {\bm S}_i\cdot{\bm S}_j\rangle_{\rm con}$ and quadrupole $C_{ij}^Q\equiv\langle {\bm Q}_i\cdot{\bm Q}_j\rangle_{\rm con}$ correlators as a function of distance between $i$- and $j$-th sites, which are shown in Fig.\,\ref{fig:correlations}\,(a)-(c). As one can see, both correlators decay exponentially in the entire SL phase, and especially $C_{ij}^Q=\frac{5}{3}C_{ij}^S$ at $\phi=0.25\pi$\,[Fig.\,\ref{fig:correlations}\,(c)] where the $SU(3)$ symmetry emerges. In addition, the dimer and chirality correlators are found to be suppressed exponentially as well. Therefore, we may conclude that there is no long-range order in this phase. 

Regarding a quantum liquid phase, the fact of existence or nonexistence of the gap is one of the most important questions. The exponential decaying of correlators in Fig.\,\ref{fig:correlations} intimates the gapped nature of the SL phase. In order to confirm this quantitatively, we propose a method for constructing the transfer matrix\,($X$):
\begin{align}
	\includegraphics[width=0.3\textwidth]{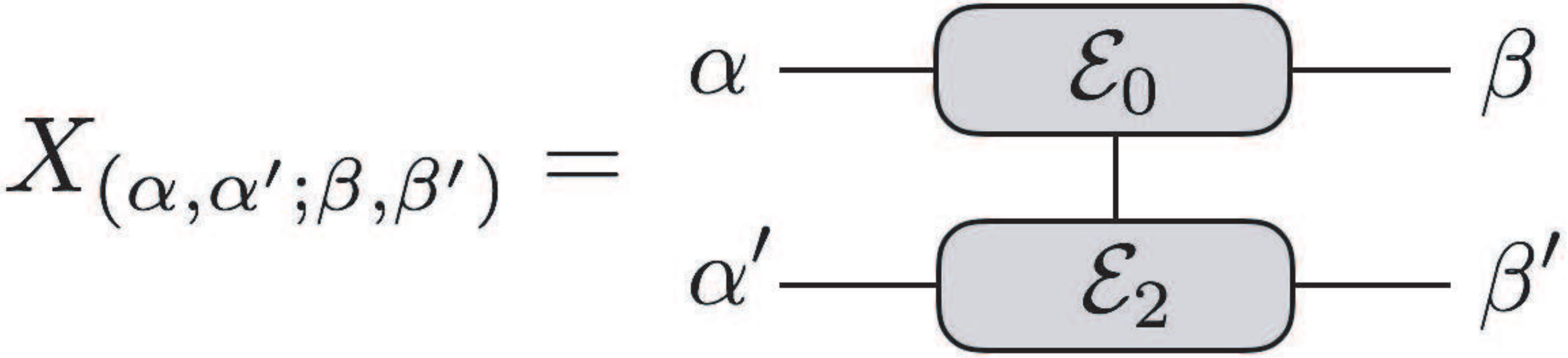},
	\label{eq:tm}
\end{align}
where $\mathcal{E}_0$ and $\mathcal{E}_2$ are the edge tensors depicted in Fig.\,\ref{fig:schematic}\,(b). Generally, the transfer matrix $X$ is supposed to contain  information on the long-range properties of the iPEPS ansatz such as the correlation length\,\cite{takahashi99}: 
%
%\begin{align}
	$\xi^{-1} = \log(\lambda_0/\lambda_1)$,
%\end{align}
%
where $\lambda_{0\,(1)}$ is the largest\,(second largest) eigenvalue of $X$. The advantage of this method of obtaining the correlation length over the one based on the measurement of two-point correlation function of certain quantities is that we do not have to know the quantity that shows the slowest decay, i.e., the present method is supposed to produce the longest correlation length that does not depend on the quantity we measure. Figure\,\ref{fig:correlations}\,(d) shows the extracted $\xi$ at $\phi=0.125\pi,0.25\pi,0.375\pi$ which converge to finite values as $D$ increases and thus clarify the existence of gap in the SL phase.

{\it Phase boundaries-} Both the boundaries of the FM phase can be fixed exactly: $\phi = -0.75\pi$ and $0.5\pi$. The boundary to the FQ phase, $\phi=-0.75\pi$, can be fixed as the point at which the system possesses the SU(3) symmetry and the FM state, and thus the FQ state can be mapped into each other by the $SU(3)$ transformation. The cusp in the energy density and discontinuities in the derivative of energy density and order parameters at $\phi=−0.75\pi$ in Fig.3 are in good agreement with the expectation and suggest the first order phase transition. At this phase boundary, the FM state from SU and the FQ state from TR-symmetric SSU come to exactly the same energy density. This is the `state-switching' phase transition at the transition point with enhanced symmetry, with the classical example being the transition of U(1) symmetric XXZ model from the easy-axis phase to the easy-plane phase at the $SU(2)$-symmetric point. The cusp and discontinuity in the energy and its derivative are also observed at the other boundary, $\phi=0.5\pi$,  suggesting the first order transition here. However, we need to note that, in contrast to the BLBQ models defined on bipartite lattices, the present system does not possess the $SU(3)$ symmetry at this phase boundary.
% because the global SU(3) transformation is constructed by different local SU(3) unitary transformation on A and B sublattices. 
Therefore, the mechanism of the transition must be somewhat different from the one at $\phi=-0.75\pi$. Nevertheless, this transition can still be located exactly. To see this, we note that there is a macroscopic GS degeneracy at $\phi=0.5\pi$. More specifically, all product states containing no $(+1,-1)$ nor $(0,0)$ nearest-neibor pairs, where $\pm 1$ or 0 are eigenvalues of $S_z$, are eigenstates of the Hamiltonian\,\cite{takahashi99}. Thus, the transition point is located exactly at $\phi=0.5\pi$ as the point where the entropy per spin becomes finite. We have obtained a rather good lower bound of the entropy per spin $S/N_s > 0.4703$ using a simple tensor network\,(see the Supplemental Material\,\cite{SM}).

%To obtain an explicit bound for the entropy density, we count the number of such GS configurations\,($Z$). Note that one can put a single 0-spin at most in a triangle plaquette, and two 0-spins in neighboring plaquettes cannot face each other. The complete $Z$ can be obtained by contracting a simple tensor network and calculating its transfer matrix\,(see the Supplemental Material\,\cite{SM}). The result for the long cylinder geometry with the system size $N = N_x N_y$ and $N_y \ll 1$ is presented in Fig.\,\ref{fig:z_xi}\,(a), where $N$ is the total number of unit-cell. The number of configurations scales $Z \sim (e^{2.822})^N=16.81^N$, and therefore $S_Z/N = \log 16.81$ is our estimate of the lower bound for the entropy density at zero temperature. 

Not only such simple product states but also some entangled states, e.g. the spin-singlet state, can be GS at $\phi=0.5\pi$. By diagonalizing the Hamiltonian of 6- and 12-site systems, we find that a spin-singlet state is degenerate at the transition point, but becomes a unique GS with a finite BL interaction. On the other hand, our $SU(2)$-symmetric iPEPS at $\phi=0.5\pi$ ansatz gives $E=1.5026$ which deviates only $0.17\%$ from the exact one $E=1.5$. Therefore, we believe reasonably that the BL interaction lifts the macroscopic degeneracy such that the singlet state gains an advantage\,(lower energy) over the product states.

\begin{figure}[!t]
  \includegraphics[width=0.5\textwidth]{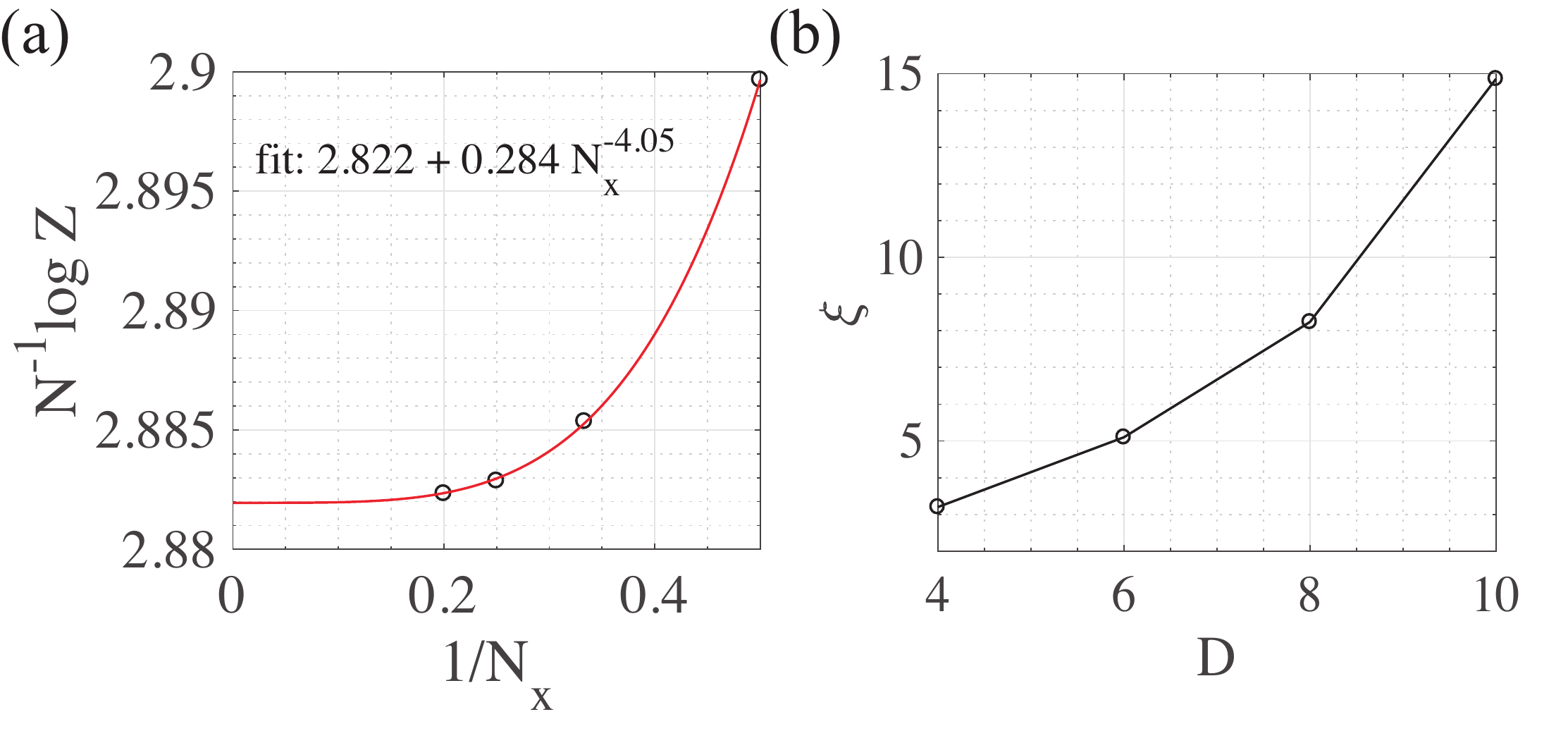}
  \caption{ (a) The number of degenerate GS at $\phi=0.5\pi$. The red solid line is the fitting curve. Here, the system is on the long cylinder with size $N=N_xN_y$ and $N_y\gg 1$. (b) Correlation length as a function of the bond dimension at $\phi=-0.25\pi$ where the phase transition between AFM and FQ occurs. }
  \label{fig:z_xi}
\end{figure}

As for the transition at $\phi\simeq -0.25\pi$, after passing it from the FQ side to the AFM side, the magnetization $M$ gradually increases from zero, and the derivative of energy density does not exhibit discontinuity around the transition\,[Fig.\,\ref{fig:EMQ}]. The correlation length is found to increase with $D$ at least up to $D=10$ as presented in Fig.\,\ref{fig:z_xi}\,(b). Those evidences strongly suggest the continuous phase transition between FQ and AFM. The nature of the transition, e.g. the universality class, might be confirmed by implementing the full update algorithm\,\cite{orus15} and tensor network renormalization methods\,\cite{gu09, evenbly15, yang17}.% to obtain a critical iPEPS ansatz and its fixed-point tensor, respectively. 

%

%The nature of transition between AFM and SL is not clear due to fact that the optimization is rather unstable in $0<\phi\lesssim 0.02$ as mentioned earlier. A significant improvement in the optimization algorithm seems to be required to overcome this uncertainty and obtain better understanding on the phase boundary, and therefore we leave it for future work. 

%
{\it Discussion-} We have explored the GS phase diagram of the $S=1$ BLBQ model on the star lattice with the state-of-art TN algorithms. In addition to FM, AFM and FQ, a gapped SL phases are identified by analyzing the local observables, various correlators and transfer matrix. In SL phase, the spin, dimer, quadrupole and chirality correlators decay exponentially, and the correlation length converges to finite even at large bond dimension of iPEPS ansatz. However, this phase has been characterized most clearly by the fact that it is represented by a PEPS with all the virtual legs of fixed parity: integer spins for inter-triangle bonds and half-integer spins for the ones forming the triangles. While the state has something in common with the Haldane phase, in which all virtual bonds having the odd parity, we are not aware of the cases where integer spins appear together with half-integer ones. Direct implication of this feature is that we may obtain different kind of boundary excitations depending on the way we cut the system. Further investigation is desirable here. The nature of phase boundaries in the model are also investigated. We confirm the first order phase transitions at $\phi=-0.75\pi$ and $0.5\pi$ by observing the phase coexistence and a finite correlation length, respectively. Particularly, the macroscopic degeneracy at $\phi=0.5\pi$ is shown by counting the partial number of degenerate GS by using the transfer matrix method\,\cite{SM}. The transition between FQ and AFM is found likely to be a continuous one due to the diverging correlation length as a function of the bond dimension. We believe that some of our work may be relevant for the star lattice antiferromagnet which has been synthesized with an iron(III) acetate hybrid material and its family\,\cite{zheng07}. We also expect to realize much of the discussed physics in optical lattices\,\cite{duan03,ruostekoski09}.
 
%Also, there are concrete and interesting open questions such as the interpretation of the level crossing in SV regarding phase transition between SL1 and SL2 or the affect of magnetic field on the phase diagram, which are remained for future work.

\acknowledgements
{\it Acknowledgements-} We would like to thank T. Okubo, R. Kaneko and S. Morita for useful discussions. The computation in the present work is executed on computers at the Supercomputer Center, ISSP, University of Tokyo, and also on K-computer (project-ID: hp170262). N.K.'s work is funded by ImPACT Program of Council for Science, Technology and Innovation (Cabinet Office, Government of Japan).	H.-Y.L. was supported by MEXT as ``Exploratory Challenge on Post-K computer"\,(Frontiers of Basic Science: Challenging the Limits).

\bibliographystyle{apsrev}
\bibliography{references.bib}

\clearpage
\onecolumngrid
\begin{center}
\textbf{\large  Supplementary Material: Spin-1 bilinear-biquadratic model on star lattice}
\end{center}

% Fixing numbering of equations/figures: Prefix 'S' and reset counter
\setcounter{equation}{0}
\setcounter{figure}{0}
\setcounter{table}{0}
\setcounter{page}{1}

\begin{center}
\parbox[t][4cm][s]{0.8\textwidth}{
	In this supplementary material, we explicitly show the macroscopic degeneracy and a lower bound for the entropy density of pure biquadratic Hamiltonian with positive coupling on the star lattice.}
\end{center}

The biquadratic Hamiltonian with positive coupling reads

\begin{align}
	H_{BQ} = \sum_{\langle i,j \rangle}	( {\bm S}_i\cdot {\bm S}_j )^2,
	\label{eq:sm_ham}
\end{align}
where $\langle i,j \rangle$ denotes the nearest-neighbor sites. Each $({\bm S}_i\cdot {\bm S}_j)^2$ operator has degenerate ground states, of which the total spin $S_{ij}^{total}$ equals to 1 and 2, with the eigenvalue $E=1$. Therefore, configurations where all neighboring spins fuse to $S_{ij}^{total}=1$ and $2$ are the ground state of Eq.\,\eqref{eq:sm_ham}. One can easily construct such states by avoiding the nearest-neighbor singlet pairs, i.e. $|+1,-1\rangle$ and $|0,0\rangle$ where $\pm1,0$ are the $S_z$ quantum number, in the whole lattice. An example on the star lattice is shown in Fig.\,\ref{fig:sm_snap_tn}\,(a). In one-dimensional chain, the total number of such configurations\,($Z$) scales $Z \simeq 2^N$ and therefore the lower bound of the entropy density is $S/N=\log 2$\,\cite{nomura91}. On the other, counting $Z$ on the star lattice is not trivial due to the loops formed by the lattice sites. We first prove the macroscopic degeneracy and extract a crude estimate for the lower bound of the entropy density by counting exactly a part of $Z$. Then, the complete $Z$ will be obtained numerically by employing a tensor network.

\section{crude estimate}

\begin{figure}[!b]
  \includegraphics[width=0.3\textwidth]{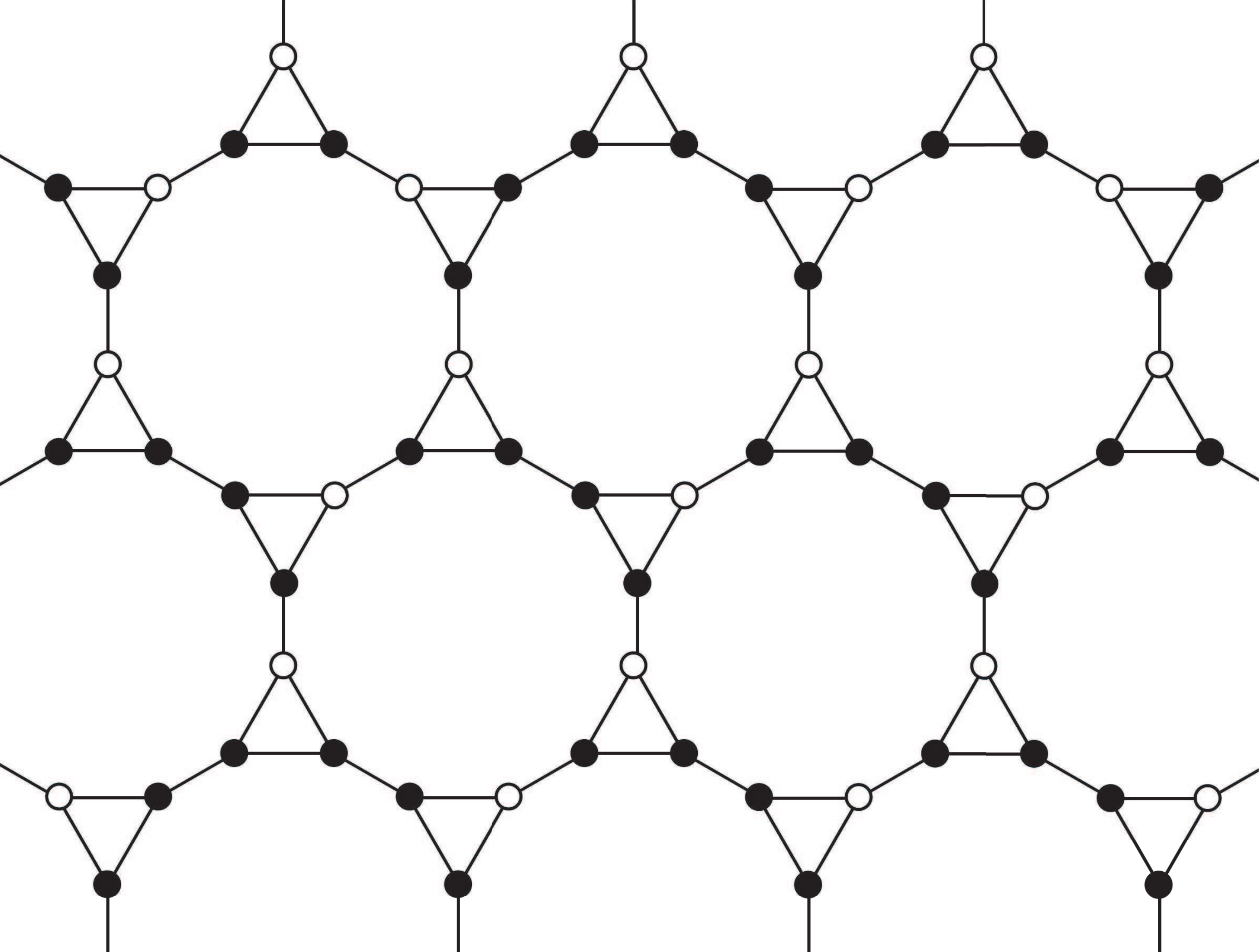}
  \caption{An exemplary ground-state configuration}
  \label{fig:sm_example}
\end{figure}

Let us consider the following configurations: spins on the upward triangles form \includegraphics[height=1em]{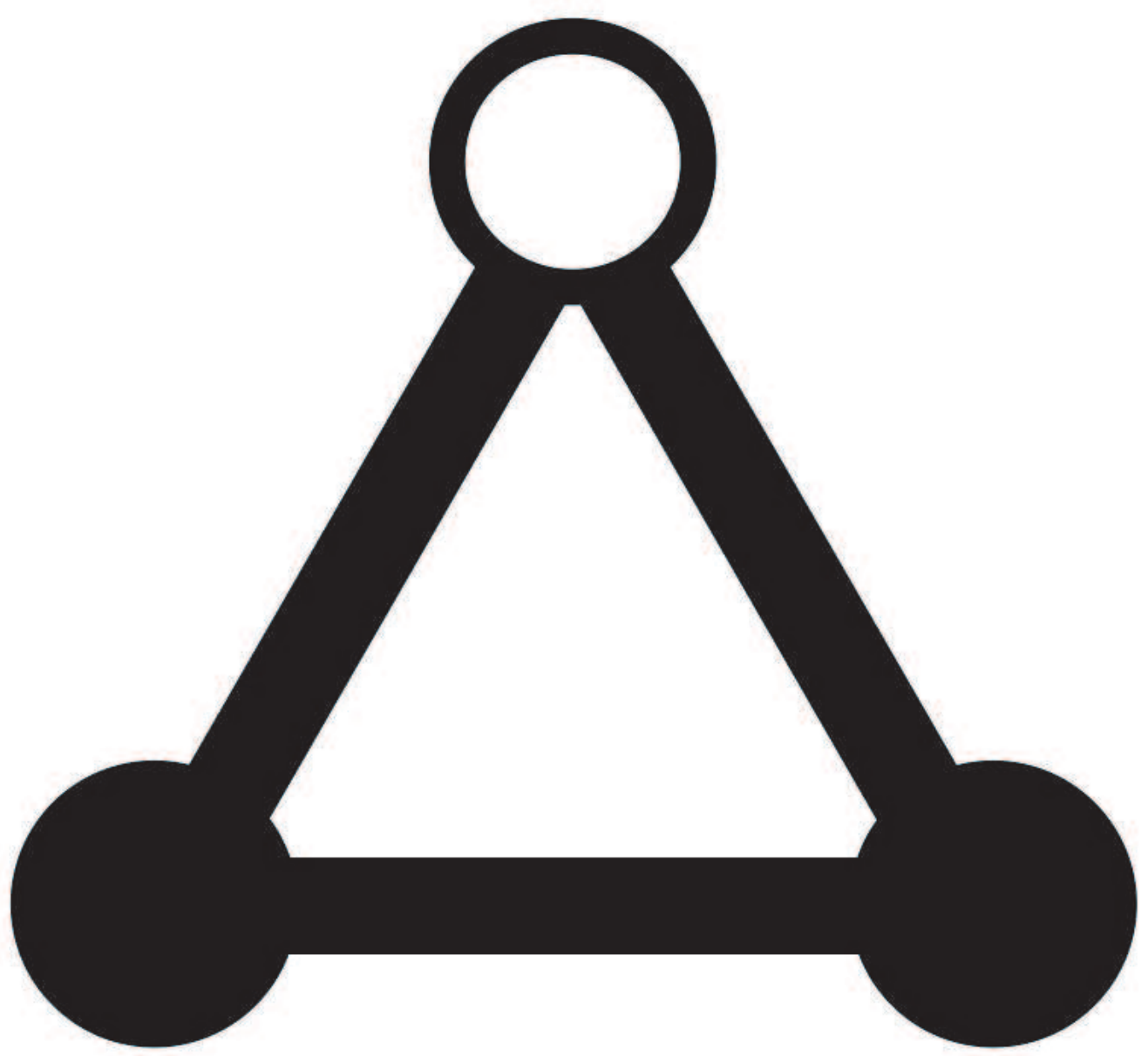} (empty: $S_z=0$, filled: $S_z=+1$ or $-1$) while \includegraphics[height=1em]{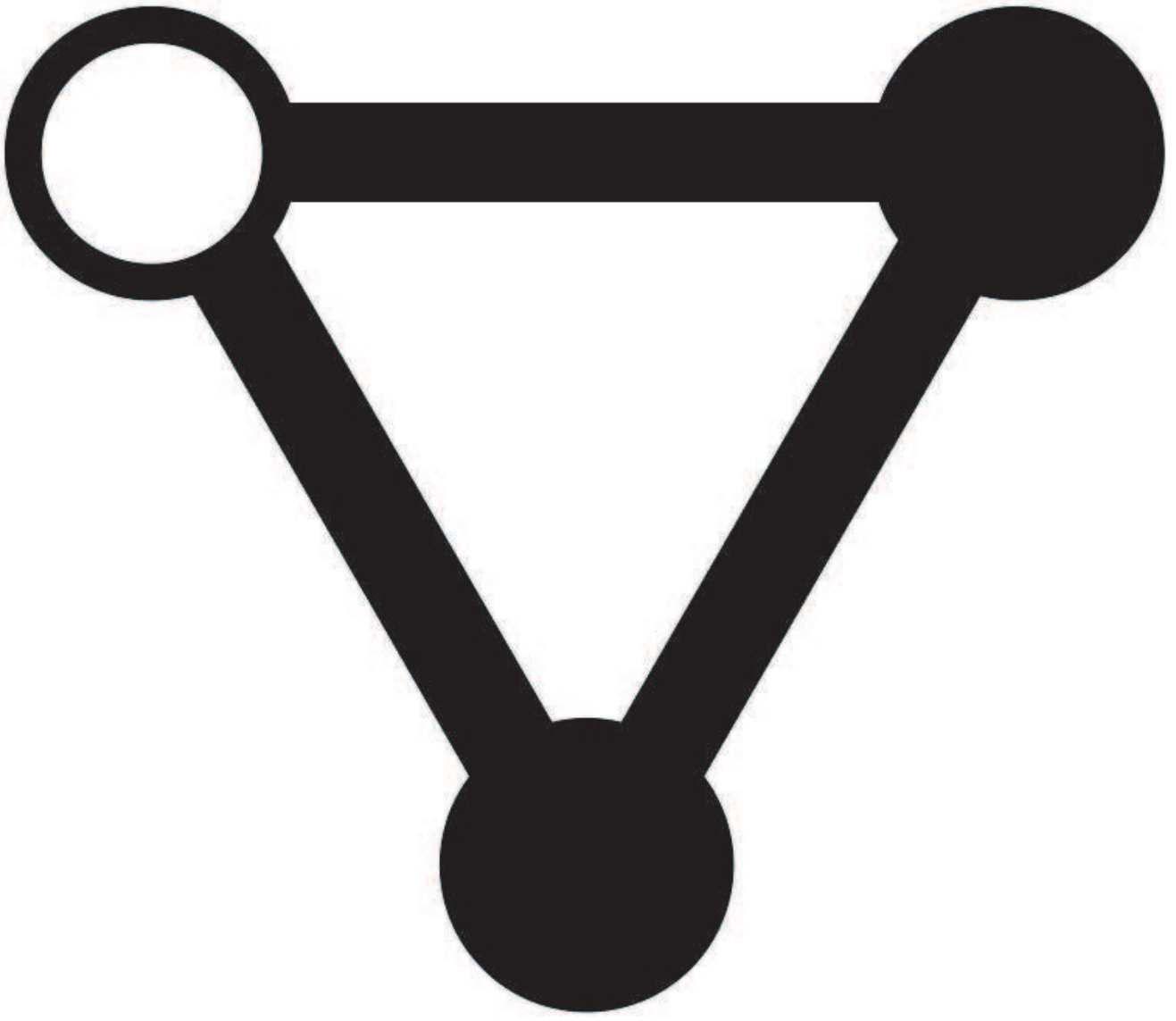} or \includegraphics[height=1em]{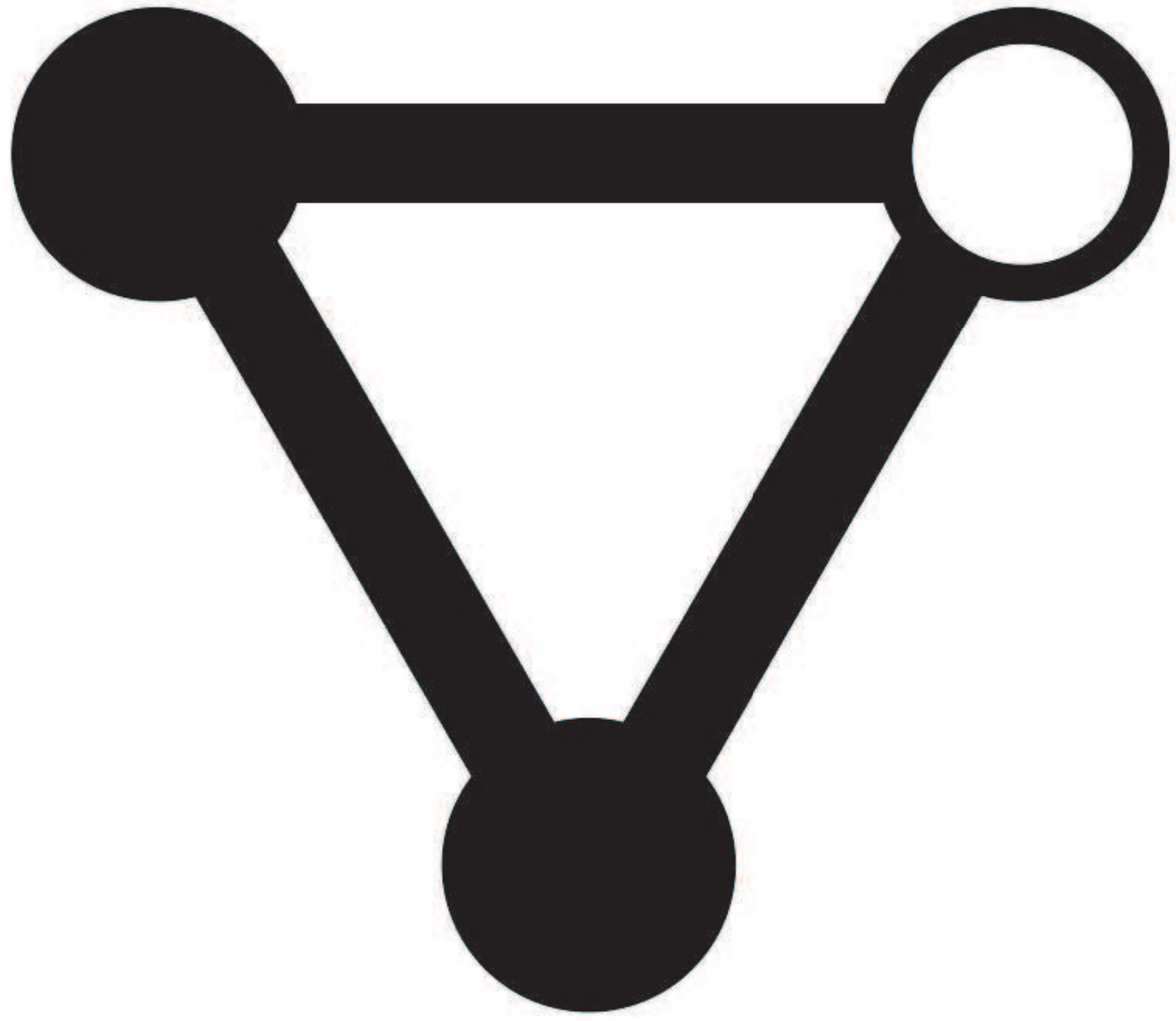} on the downward triangles. An example is shown in Fig.\,\ref{fig:sm_example}.
The number of such configurations is simply $Z_0 = 2^{N_{\vartriangle}+N_{\triangledown}}$ where $N_{\vartriangle\,(\triangledown)}$ is the number of upward\,(downward) triangle plaquettes. The factor $2^{N_{\vartriangle}}$ comes from the fact that there are $N_{\vartriangle}$ clusters bounded by the 0-spins, whereas $2^{N_{\triangledown}}$ from the choice between \includegraphics[height=1em]{dt1.eps} or \includegraphics[height=1em]{dt2.eps} for each downward triangle. Consequently, the entropy density is $S_{Z_0}/N = \log 4$, where $N$ is the total number of unit-cell.

\section{tensor network representation for counting $Z$}

One can map counting $Z$ into a dimer packing problem with a particular constraint. The basic idea is the following. One can regard the neighboring pair $|0,\pm1\rangle$ connecting two triangles as a directed\,(from 0 to $\pm1$) dimer, e.g.
\begin{align}
	\includegraphics[width=0.4\textwidth]{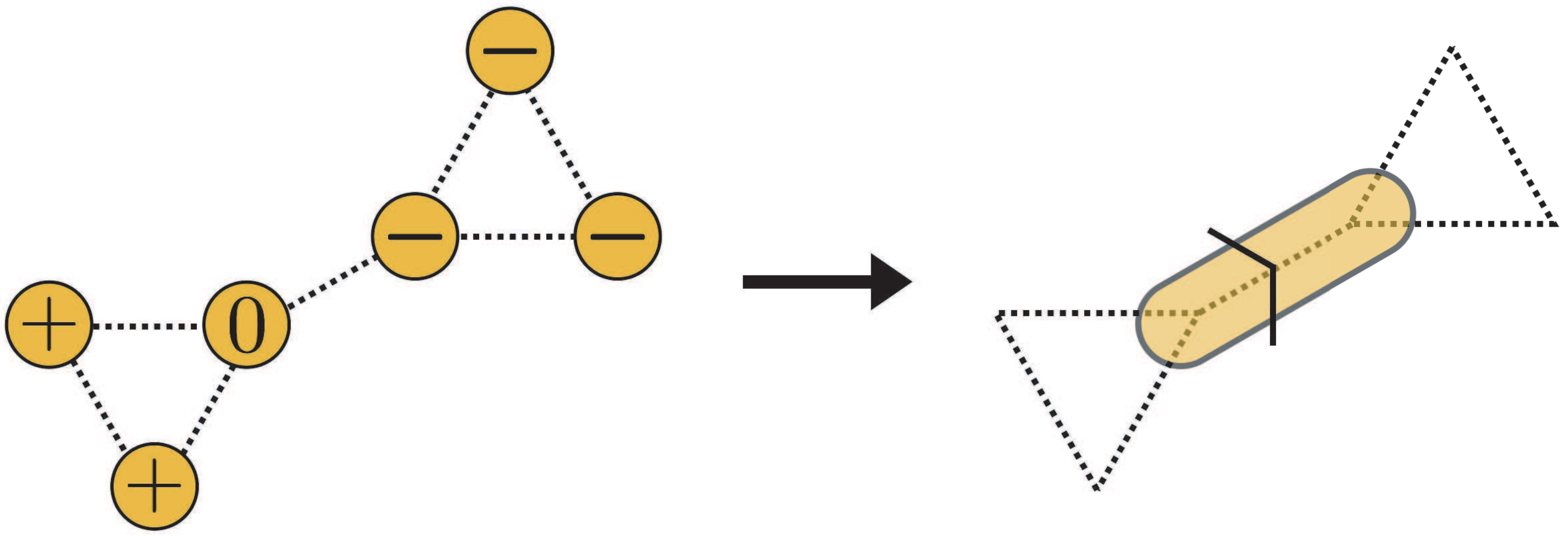}.
	\nonumber
	\label{eq:sm_dimer}
\end{align}
\begin{figure}[!h]
  \includegraphics[width=0.5\textwidth]{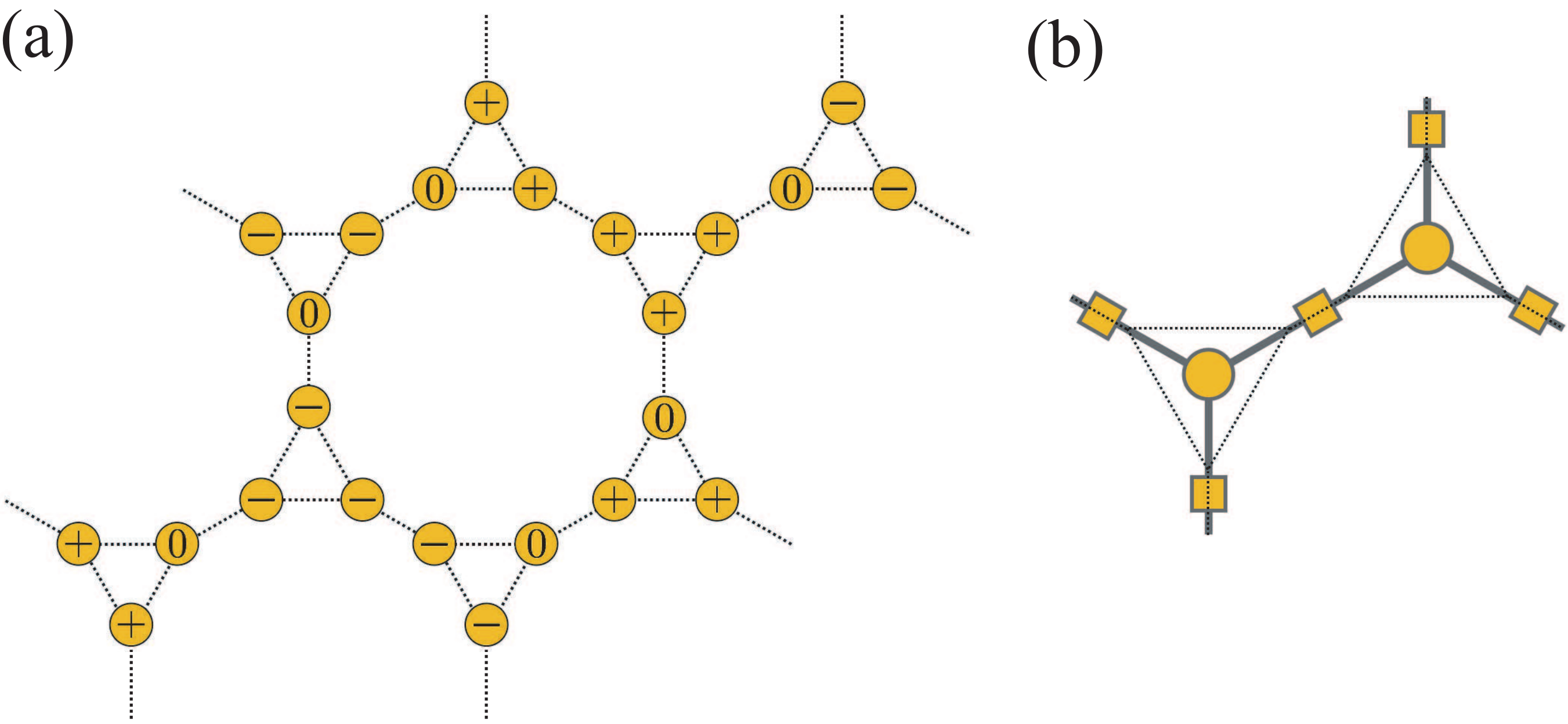}
  \caption{(a) An exemplary ground state configuration of the Hamiltonian in Eq.\,\eqref{eq:sm_ham}, and (b) a proposed tensor network, made of plaquette and bond tensors, counting such configurations. }
  \label{fig:sm_snap_tn}
\end{figure}
Here, the direction is necessary to distinguish the configuration $|0,\pm1\rangle$ and $|\pm1,0\rangle$. Therefore, we assign four species of dimers on every bond and then count the number of all dimer configurations allowing the hole with an equal weight. Additional constraint is that only a single outgoing dimer is permitted on every triangle plaquette at most. Now, one can count the total number of such dimer configurations by employing a simple tensor network composed of rank-3 plaquette tensors\,($P_{ijk}$) on the center of triangle loop and bond matrices\,($B_{ij}$) connecting $P$ tensors as depicted in Fig.\,\ref{fig:sm_snap_tn}\,(b). For simplicity, let us first consider the configurations with only $S_z = 0, +1$ excluding $S_z=-1$ state. Then, the bond dimension $D=3$ is required, and each state on the leg can be defined as follows: 

\begin{itemize}
	\item $|0\rangle$: start-point of dimer on the vertex
	\item $|1\rangle$: end-point of dimer on the vertex
	\item $|2\rangle$: hole on the vertex
\end{itemize}
Due to the constraint allowing only a single outgoing dimer at most on each triangle plaquette, the configurations in Fig.\,\ref{fig:sm_tensor} and their cyclic permutation partners are only non-zero elements of the tensor $P_{ijk}$ while the bond matrix is
\begin{align}
	B = \begin{pmatrix}
		0 & 1 & 0\\ 1 & 0 & 0 \\ 0 & 0 & 1
	\end{pmatrix}.
\end{align}
\begin{figure}[!h]
  \includegraphics[width=0.5\textwidth]{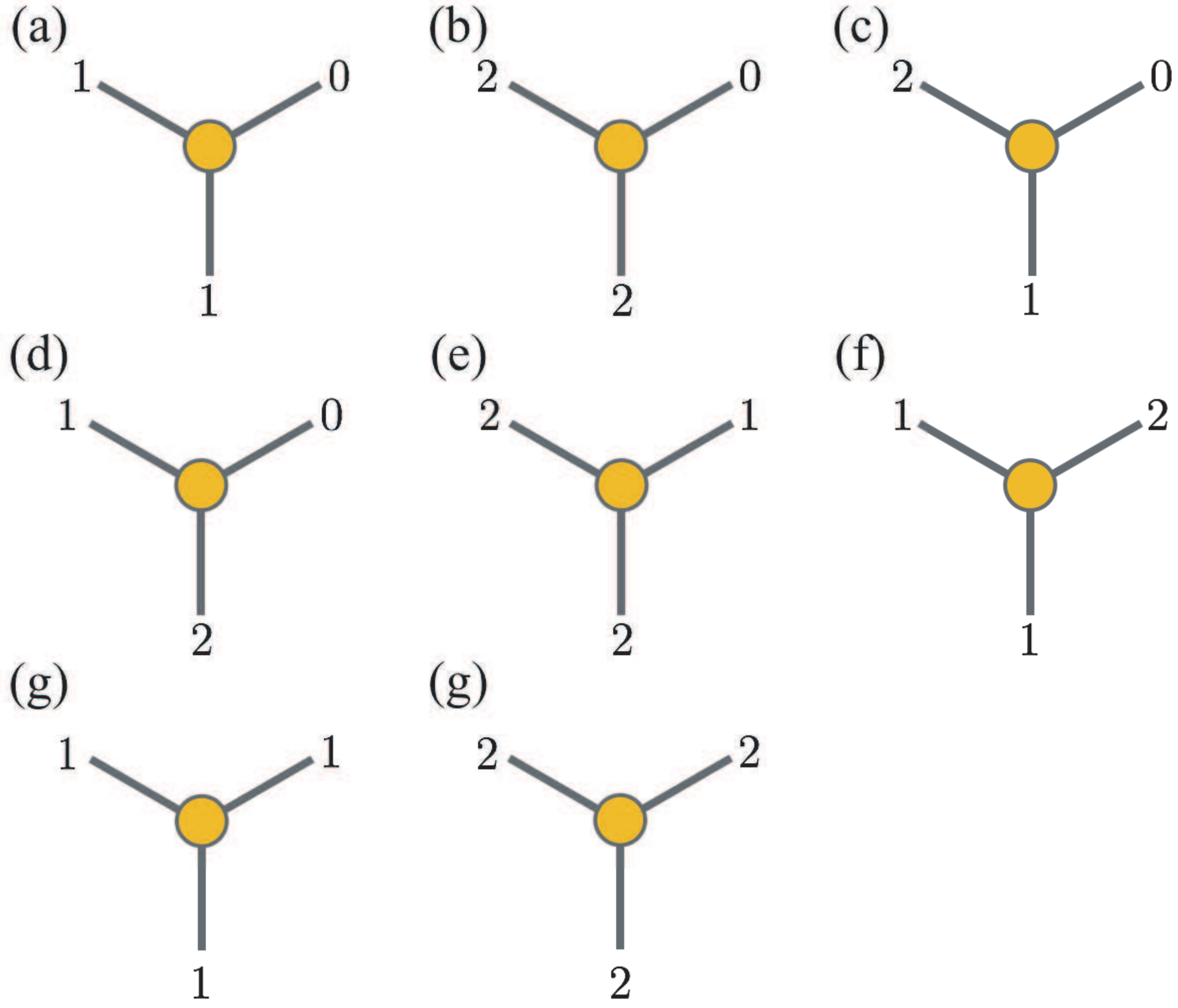}
  \caption{Non-zero elements of the plaquette tensor $P_{ijk}$ to count the configurations with only $S_z=0,+1$ states.}
  \label{fig:sm_tensor}
\end{figure}
We set the wight of each configuration in Fig.\,\ref{fig:sm_tensor} to be 1, and then the contraction of the tensor network give the total number of the ground state made of only $S_z = 0,+1$ states. Similarly, one may count the configurations even including $S_z=-1$ state by enlarging the bond dimension to $D=5$. 

Now, we contract two plaquette tensors and three bond tensors to have a translational invariant tensor network on the square lattice. The efficient calculation of $Z$ proceeds on a cylinder geometry with the periodic boundary condition imposed along the $x$-direction of length $N_x$ and open ends along the $y$-direction of length $N_y$. By contracting the tensors along the $x$-direction, one obtains the so-called row-to-row transfer matrix as schematically depicted below
\begin{align}
	\includegraphics[width=0.25\textwidth]{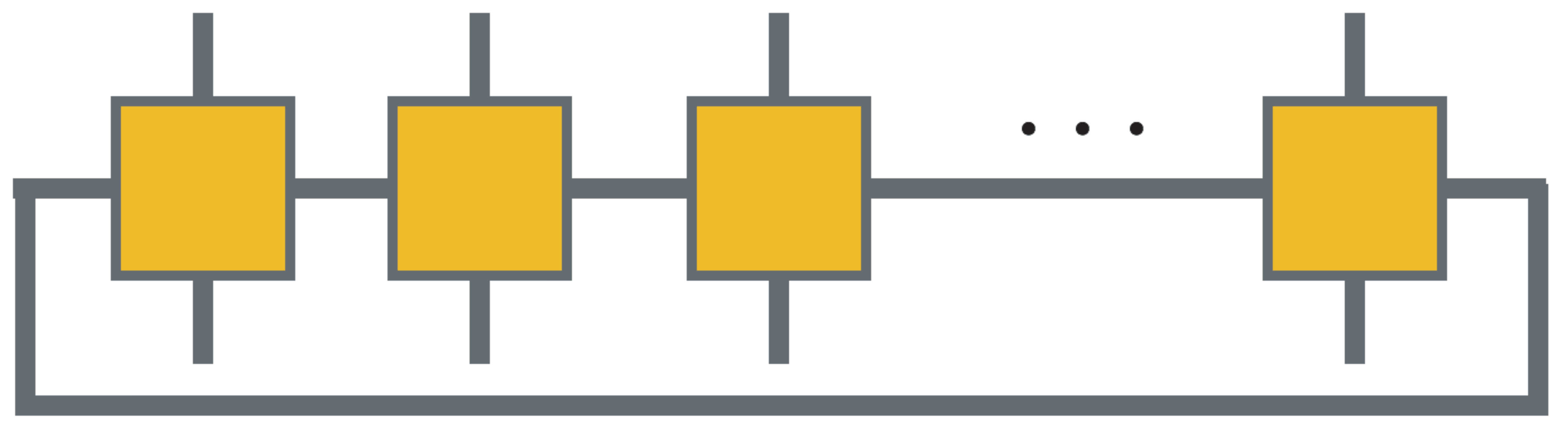}
	\nonumber
\end{align}
\begin{figure}[!t]
  \includegraphics[width=0.35\textwidth]{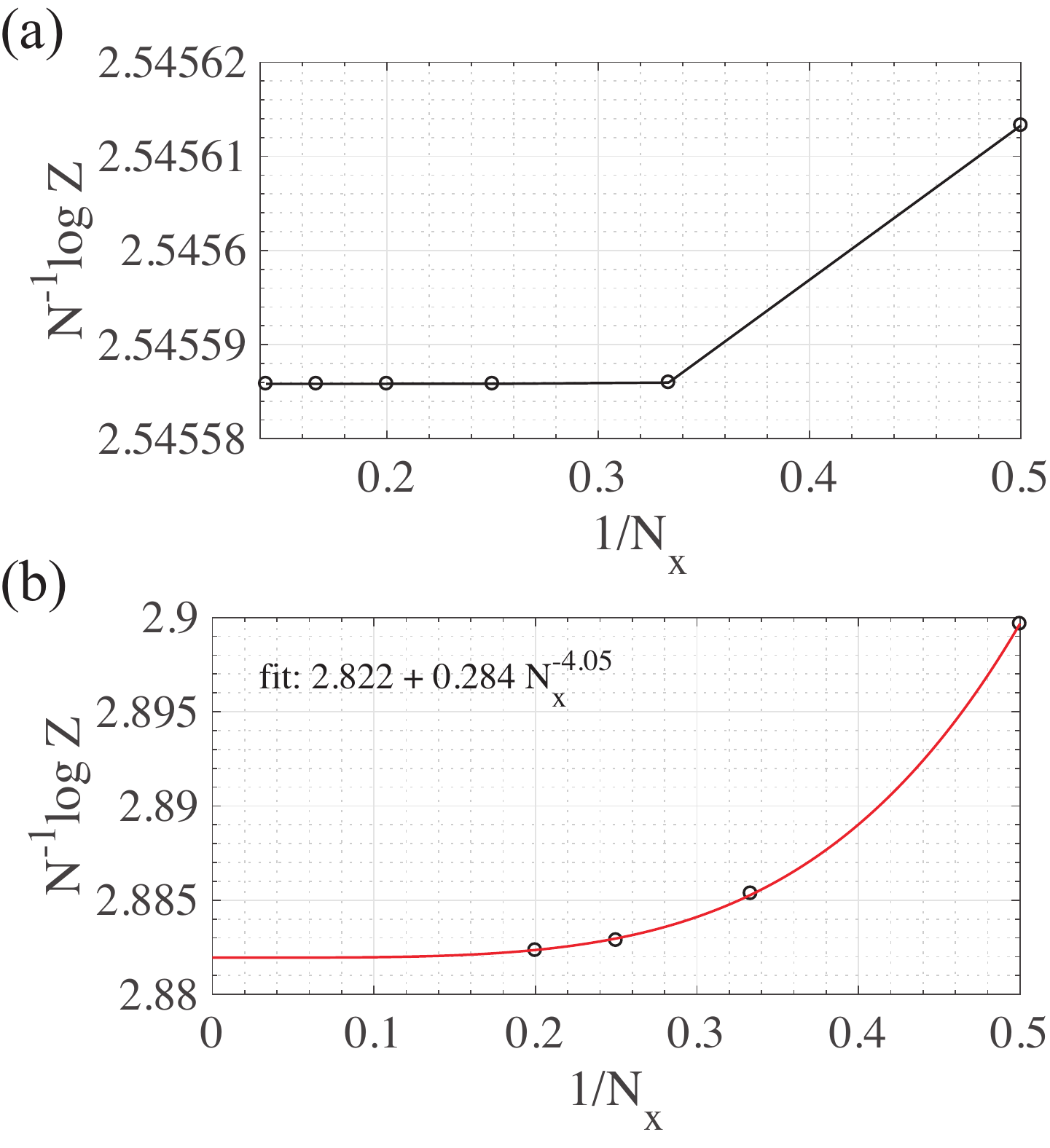}
  \caption{ The scaling of the number of GS configurations made of (a) $S_z = 0,+1$ (b) $S_z = 0,\pm1$, respectively.}
  \label{fig:sm_Z}
\end{figure}
Assuming $N_y\gg1$, the total number of configurations $Z$ scales like $Z_{N_x}\simeq (\lambda_{N_x})^{N_y}$ where $\lambda_{N_x}$ is the largest eigenvalue of the transfer matrix with a length $N_x$. Even though the transfer matrix is not Hermitian\,(or symmetric), but the largest eigenvalue is unique and real by the Perron-Frobenius theorem\,\cite{horn90}. In order to obtain the scaling behavior of $Z$ in terms of the system size $N=N_xN_y$, we plot $N^{-1}\log Z_{N_x} = N_x^{-1}\log\lambda_{N_x}$. The result is presented in Fig.\,\ref{fig:sm_Z}\,(a). As one can see, the entropy density with $N_x = 3$ is already very close to the one in the thermodynamic limit, which is $N^{-1}\log Z_{N_x=\infty}\simeq 2.54587$. We therefore conclude that the number of degenerate GS made of only $S_z = 0,+1$ states scales $Z \sim (e^{2.545587})^N = 12.75^N $. 

By enlarging the bond dimension to $D=5$, one can evaluate the degeneracy including $S_z=-1$ configurations. It is easy to find the non-zero elements of tensor, which is straightforward extension from the elements in Fig.\,\ref{fig:sm_tensor}. The entropy density is shown in Fig.\,\ref{fig:sm_Z}\,(b). Here, we extrapolate the data to obtain the one in the thermodynamic limit. Now, the number of degenerate states scales $Z \sim (e^{2.822})^N=16.81^N$, and this can be regarded as the lower bound for the true entropy density of pure biquadratic model on the star lattice.

\end{document}